\documentclass[english,final,5p,times,twocolumn,sort&compress,superscriptaddress]{elsarticle}
\usepackage[T1]{fontenc}
\usepackage{color}
\usepackage{babel}
\usepackage{amsmath}
\usepackage{amstext}
\usepackage{graphicx}
\usepackage{esint}
\usepackage[unicode=true,
 bookmarks=true,bookmarksnumbered=false,bookmarksopen=false,
 breaklinks=false,pdfborder={0 0 0},pdfborderstyle={},backref=false,colorlinks=true]
 {hyperref}
\hypersetup{
 pdfauthor={Ambroise van Roekeghem}}

\makeatletter
\@ifundefined{date}{}{\date{}}
\journal{Computer Physics Communications}
\date{\today}
\usepackage{cuted}

\makeatother

\begin{document}

\begin{frontmatter}{}

\title{Quantum Self-Consistent Ab-Initio Lattice Dynamics}

\author[cea]{Ambroise van Roekeghem\corref{cor1}} 
\author[tuw]{Jes\'us Carrete}
\author[cea]{Natalio Mingo}
\cortext[cor1]{Corresponding author.\\\textit{E-mail address:} ambroise.vanroekeghem@cea.fr}
\address[cea]{Universit\'{e} Grenoble Alpes, CEA, LITEN, 17 rue des Martyrs, 38054 Grenoble, France}
\address[tuw]{Institute of Materials Chemistry, TU Wien, A-1060 Vienna, Austria}
\begin{abstract}
The Quantum Self-Consistent Ab-Initio Lattice Dynamics package (QSCAILD) is a python library that computes temperature-dependent effective 2nd and 3rd order interatomic force constants in crystals, including anharmonic effects. QSCAILD's approach is based on the quantum statistics of a harmonic model. The program  requires the forces acting on displaced atoms of a solid as an input, which can be obtained from an external code based on density functional theory, or any other calculator. This article describes QSCAILD's implementation, clarifies its connections to other methods, and illustrates its use in the case of the SrTiO$_{3}$ cubic perovskite structure.
\end{abstract}

\end{frontmatter}{}

\noindent
{\bf PROGRAM SUMMARY}

\begin{small}
\noindent      
{\em Program Title:} QSCAILD\\           
{\em Licensing provisions:} GNU General Public License version 3.0\\       
{\em Programming language:} Python\\
{\em External routines/libraries:} MPI, NumPy, SciPy, spglib, phonopy, sklearn\\      
{\em Nature of problem:} Calculation of effective interatomic force constants at finite temperature\\
{\em Solution method:} Regression analysis of forces from density functional theory coupled with a harmonic model of the quantum canonical ensemble, performed in an iterative way to achieve self-consistency of the phonon spectrum\\     
\end{small}

\section{Introduction}

\label{sec:Introduction}

Lattice anharmonicity is at the origin of basic phenomena in solid state physics, such as thermal expansion, displacive phase transitions or intrinsic thermal resistivity. In the quasi-harmonic approximation, the volume dependence of lattice vibrations is taken into account to model thermal expansion and as a consequence the phonon spectrum becomes temperature-dependent. However, the definition of phonons in an anharmonic potential as interacting quasiparticles with a temperature-dependent population also implies that even at constant volume their energy can depend on temperature. The last decade has seen a tremendous improvement
in the quantitative description of such temperature-dependent phonon properties in solids based on Density Functional Theory.
A precursor of current methods is the Self-Consistent Ab-Initio
Lattice Dynamics (SCAILD) approach proposed by Souvatzis et al. in
2008 \citep{Souvatzis_SCAILD}, which is based on a self-consistent
calculation of phonon frequencies using a collection of supercells
in which atoms were displaced according to the thermal mean square
displacement in the classical limit. Subsequent developments are the Temperature Dependent Effective Potential (TDEP) method \citep{Hellman_TDEP_2011}, based
on fitting effective force constants to the results of ab initio molecular
dynamics;
and the Stochastic Self-Consistent Harmonic Approximation (SSCHA) \citep{Errea_SSCHA}, which aims at mimizing the free
energy of the system within a harmonic density matrix ansatz. More
recently, other methods and implementations have appeared (see for instance Refs.\ \citep{Tadano_SCPH,Xia_APRN,Carreras_DynaPhoPy,Navaneetha_anharmonic,hiphive}), showing that the field is becoming broader and raising increasing
interest. The approach that we describe in this paper was first implemented
in 2016 \citep{Ambroise_ScF3} and has been refined and improved since then. A similar procedure is also described in Ref.\ \citep{Shulumba_PbSe}. It is directly related to the original SCAILD approach, the main difference being that it lifts the approximation to the statistics of atomic displacements made at the time.
This is why we name it QSCAILD, for Quantum
Self-Consistent Ab-Initio Lattice Dynamics. Below we describe
the technical details of the implementation that we publicly release
and show a selection of results obtained with the code.

\section{Methodology}

\subsection{Basic principle}
For $N$ ions with a harmonic Hamiltonian $\mathcal{H}$, let $\rho_{\mathcal{H}}\left(u\right)$ be the probability of finding the system in
a configuration in which each ion $i$ is displaced in Cartesian direction $\alpha$
by $u_{i\alpha}$. This probability is proportional, up to a normalization factor, to $\exp\left(-\frac{1}{2}u^{T}\mathbf{\Sigma}^{-1}u\right)$,
where $\Sigma\left(i\alpha,j\beta\right)$ is the quantum covariance for
atoms $i,j$ and directions $\alpha,\beta$:\citep{Ambroise_ScF3} 
\[
\Sigma\left(i\alpha,j\beta\right)=\frac{\hbar}{2\sqrt{M_{i}M_{j}}}\sum_{m}\omega_{m}^{-1}\left(1+2n_{B}\left(\omega_{m}; T\right)\right)\epsilon_{mi\alpha}\epsilon_{mj\beta}^{*}.
\]

\begin{figure}
\begin{centering}
\includegraphics[width=8cm]{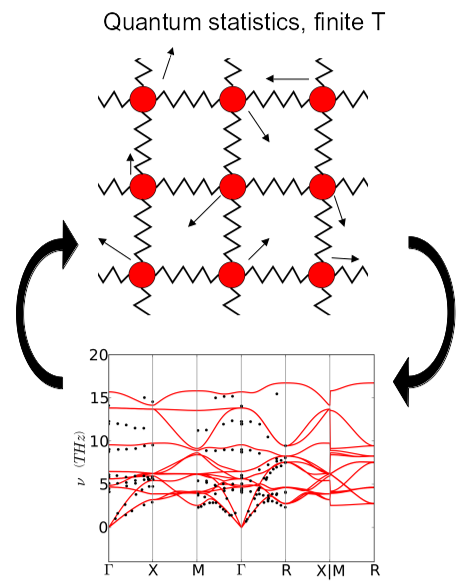}
\par\end{centering}
\caption{Simplified sketch of the QSCAILD method.\label{QSCAILD}}

\end{figure}

\noindent Here, $M_i$ is the $i$-th atomic mass, $\omega_{m}$ the phonon frequency
of mode $m$ (comprising both wavevector and branch degrees of freedom),
$\mathbf{\epsilon}_{m}$ the corresponding eigenvector and $n_{B}$
the Bose-Einstein distribution at temperature $T$. The method obtains a self-consistent set of interatomic force constants by the following iterative procedure:
\begin{itemize}
    \item The matrix $\mathbf{\Sigma}$ is computed from the phonon frequencies and eigenvectors and used to generate a random set of atomic displacements $\left\{ u_{i\alpha}^{(n)}\right\} $, with $i\alpha$ indexing each atom and direction in the supercell and $n$ denoting each given random configuration.
    \item The forces $\left\{f_{i\alpha}^{(n)}\right\}$ acting on each atom of the supercell generated by this set of displacements $\left\{ u_{i\alpha}^{(n)}\right\} $ are used to fit the interatomic force constants of a model potential, using least-squares minimization.
\end{itemize}
This scheme is depicted in Fig.~\ref{QSCAILD}.

\subsection{General workflow}

\begin{figure}
\begin{centering}
\includegraphics[width=8cm]{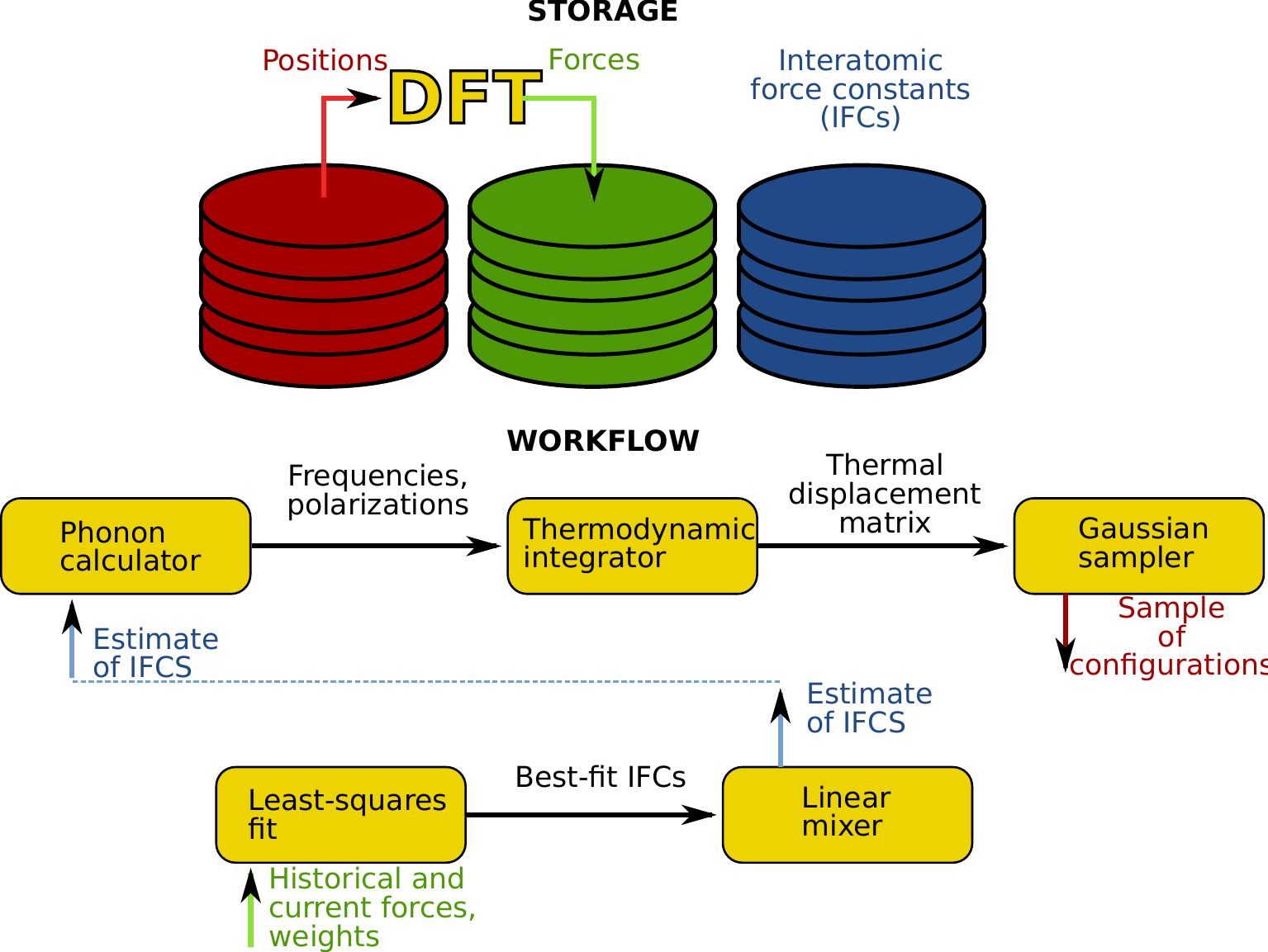}
\par\end{centering}
\caption{Workflow of the published implementation. This workflow can be wrapped in a driver that takes care of the optimization of lattice parameters and atomic positions.\label{QSCAILD_code}}

\end{figure}

The code is a Python library that can perform two tasks:
\begin{enumerate}
\item Starting from a set of interatomic force constants and a crystal structure, compute the thermal displacement matrix and generate a number of displaced configurations.
\item After the forces on the atoms in those configurations have been computed from DFT, collect the results and generate new effective force constants.
\end{enumerate}
Between those two tasks, the user has to perform the DFT calculations
using a workflow that is system-dependent and is thus not implemented
by the code. This scheme has to be iterated until convergence
is reached. The program keeps track of the successive iterations and can monitor the convergence. At present, the code is written to use input from the
Vienna Ab-initio Simulation Package \citep{VASP1,VASP2} and Phonopy \citep{phonopy}
force constants in full format, but it can be adapted to
use another DFT code. It also uses a modified version of the thirdorder
script \citep{ShengBTE_2014}. The output is natively compatible with the ShengBTE
\citep{ShengBTE_2014} and almaBTE \citep{almaBTE_CompPhysComm2017}
software packages, so that thermal conductivity calculations including anharmonic
renormalization of force constants can be performed in a straightforward
way.
This workflow is illustrated in Fig.~\ref{QSCAILD_code}.

\subsection{Links to other methods}

The link between the different methods cited in the introduction has
not been discussed in depth in the literature, in particular regarding the
connection between methods based on fitting forces from AIMD and methods based on minimizing the free energy within a harmonic Hilbert space. Here we show that
the solution of the QSCAILD method is equivalent to the solution of
the SSCHA, and thus holds the same properties, notably the minimization
of the free energy. For each iteration, a sample of configurations
is generated with density matrix $\rho_{\mathcal{H}}$. The forces
are calculated and fitted to a harmonic force constant matrix $\Phi$
by least squares minimization of the quantity $S=\sum_{k,n}\left(s_{k}^{(n)}\right)^{2}$  over all sets of force constants compatible with the symmetries,
with $s_{k}^{(n)}=f_{k}^{(n)}+\sum_{l}\Phi_{kl}u_{l}^{(n)}$, in which
$n$ is a given configuration, and $k,l$ reduced indices for atom
and Cartesian coordinates. The solution of the least squares problem
satisfies  $\frac{\partial S}{\partial\Phi_{ij}}=0$ for each $i,j$, so

\begin{eqnarray*}
2\sum_{k,n}s_{k}^{(n)}\frac{\partial s_{k}^{(n)}}{\partial\Phi_{ij}} & = & 0\\
\Rightarrow 2\sum_{k,n}s_{k}^{(n)}\frac{\partial\sum_{l}\Phi_{kl}u_{l}^{(n)}}{\partial\Phi_{ij}} & = & 0\\
\Rightarrow \sum_{n}s_{i}^{(n)}u_{j}^{(n)} & = & 0\\
\Rightarrow
\sum_{n}\left(f_{i}^{(n)}-f_{\mathcal{\mathcal{H}}^{new}i}^{(n)}\right)u_{j}^{(n)} & = & 0.
\end{eqnarray*}

Here, $\mathcal{H}^{new}$ is the Hamiltonian of the next iteration
and $f_{\mathcal{H}^{new}k}=-\sum_{l}\Phi_{kl}u_{l}^{(n)}$ is the
harmonic force. In the limit of an infinite number of configurations,
we have for all $i,j$:

\begin{equation}
\int_{\mathbb{R}^{3N}}\left[\left(f_{i}-f_{\mathcal{H}^{new}i}\right)u_{j}\right]\rho_{\mathcal{H}}(u)du=0\label{eq:least-squares}
\end{equation}

When self-consistency is achieved, $\mathcal{H}^{new}=\mathcal{H}$,
and the gradient of the free energy with respect to force constants
coefficients in the sense of the SSCHA is zero, according to equation
(22) of Ref.\ \citep{Errea_PRB_SSCHA}. Together with equation (17) of Ref.\ \citep{SSCHA_free_energy}, this shows that, at fixed point of the iteration (i.e., if and when convergence is achieved), the least-square fitting of forces is equivalent to performing the average of second derivatives of the potential. It is thus also equivalent
to the approach described in Ref.\ \citep{Tadano_SCPH}. A similar
fit can be done using AIMD atomic configurations, in which case
the harmonic density matrix is replaced by the MD density matrix,
which does not contain zero point motion, but in turn is not limited to
a harmonic form. In the case of a non-least-squares fit such as what
is done with the LASSO algorithm, the minimization of the free energy
is balanced with the requirement to obtain a sparse set of force constants \citep{Zhou_compressive_sensing}.
Finally, we note that including the 3rd order force constants in the
least-squares equation does not change the self-consistent solution:
 since $s_{k}^{(n)}=f_{k}^{(n)}+\sum_{l}\Phi_{kl}u_{l}^{(n)}+\sum_{lm}\Psi_{klm}u_{l}^{(n)}u_{m}^{(n)}$,
 we now have 

\begin{eqnarray*}
\sum_{k,n}s_{k}^{(n)}\frac{\partial s_{k}^{(n)}}{\partial\Phi_{ij}}=\sum_{n}r_{i}^{(n)}u_{j}^{(n)} & = & 0\\
\Rightarrow \sum_{n}\left(f_{i}^{(n)}-f_{\mathcal{\mathcal{H}}^{new}i}^{(n)}\right)u_{j}^{(n)}+\sum_{lm,n}\Psi_{ilm}u_{l}^{(n)}u_{m}^{(n)}u_{j}^{(n)} & = & 0.
\end{eqnarray*}

\noindent As we are working with a harmonic density matrix, for each
$l,m,j$ we have
\begin{equation}
\int_{\mathbb{R}^{3N}}\left(u_{l}u_{m}u_{j}\right)\rho_{\mathcal{H}}(u)du=0\label{eq:parity}
\end{equation}
and the condition \eqref{eq:least-squares} is recovered. For the third
order part, we expand:

\begin{eqnarray*}
\sum_{k,n}s_{k}^{(n)}\frac{\partial s_{k}^{(n)}}{\partial\Psi_{hij}}=\sum_{n}s_{h}^{(n)}u_{i}^{(n)}u_{j}^{(n)} & = & 0,\\
\sum\limits_{n}\left(f_{h}^{(n)}+\sum\limits_{lm}\Psi_{hlm}u_{l}^{(n)}u_{m}^{(n)}\right)u_{i}^{(n)}u_{j}^{(n)}&&\\+\sum\limits_{l,n}\Phi_{hl}u_{l}^{(n)}u_{i}^{(n)}u_{j}^{(n)}&=&0.
\end{eqnarray*}

\noindent and, similarly, the second term on the left-hand side can be dropped
in the limit of an infinite number of configurations, so the above equation
becomes:

\begin{equation}
\int_{\mathbb{R}^{3N}}\left[\left(f_{h}-f_{\mathcal{H}^{new}h}^{(3rd)}\right)u_{i}u_{j}\right]\rho_{\mathcal{H}}(u)du=0
\end{equation}

More generally, and for the same reasons related to the symmetry of the density matrix, an expansion of $f_{i}$
to all orders shows that, for fixed atomic positions, only even orders can renormalize the second-order part while odd orders renormalize the third order part of the effective Hamiltonian. When average atomic positions are not fixed by symmetry, their modification due to odd-order anharmonicity can have an impact on second order effective force constants. These new average positions can be computed by minimizing the average force, as described in Eq.\ (21) of Ref.\ \citep{Errea_PRB_SSCHA}. A sample implementation is included in our code, although more complex workflows require the user to write a more complete driver. We also stress that here the density matrix is computed solely based on the second-order force constants to stay harmonic, although in principle this sampling could be improved.

In spite of equation \eqref{eq:parity} showing that including only second-order or both second and third-order terms in the fit should yield the same self-consistent effective harmonic force constants in the limit of an infinite number of configurations, in practice, and due to the necessarily finite number of samples, the inclusion of both second- and third-order terms improves the stability and speed of convergence of the algorithm. This result is not necessarily intuitive since the number of unknowns in the system is clearly increased. In general, we recommend a strongly overdetermined system with $5$ to $10$ times more forces than irreducible unknowns.\newline

\subsection{Potential and kinetic pressure}

Below we show how we include the kinetic pressure term in our static
simulation and the relation with the quasiharmonic approximation and
with molecular dynamics. 

A quantity that can be obtained from the DFT calculations is the stress tensor for each configuration, which
contains only the derivative of the potential energy. We obtain the
value of the diagonal elements $\{u\}$ in the case of a potential with
2nd and 3rd order terms, and we consider only the isotropic
case for simplicity. When the volume changes from $V$ to
$V+\Delta V$, the positions of the atoms in the cell change from
$r_{i}=r_{i}^{eq}+u_{i}$ to $r_{i}\left(1+\frac{\Delta V}{3V}\right)$,
such that the displacement with respect to equilibrium $u_{i}$ becomes
$u_{i}+\Delta_{i}$ with $\Delta_{i}=r_{i}^{eq}\frac{\Delta V}{3V}+u_{i}\frac{\Delta V}{3V}$.
We thus obtain, to the first order:\begin{strip}

\[
\Delta E_{p}=\frac{1}{2}\sum_{ij}\Phi_{ij}\left(\Delta_{i}u_{j}+u_{i}\Delta_{j}\right)+\frac{1}{6}\sum_{ijk}\Psi_{ijk}\left(\Delta_{i}u_{j}u_{k}+u_{i}\Delta_{j}u_{k}+u_{i}u_{j}\Delta_{k}\right)
\]

\[
\frac{\Delta E_{p}}{\Delta V}=\frac{1}{6V}\sum_{ij}\Phi_{ij}\left(r_{i}^{eq}u_{j}+u_{i}r_{j}^{eq}\right)+\frac{1}{6V}\left(2\sum_{ij}\Phi_{ij}u_{i}u_{j}\right)+\frac{1}{18V}\sum_{ijk}\Psi_{ijk}\left(r_{i}^{eq}u_{j}u_{k}+u_{i}r_{j}^{eq}u_{k}+u_{i}u_{j}r_{k}^{eq}\right)+\frac{1}{18V}\left(3\sum_{ijk}\Psi_{ijk}u_{i}u_{j}u_{k}\right).
\]

Thus, the average of the diagonal elements of the stress tensor in
the limit of infinite number of configurations is:

\begin{equation}
\int_{\mathbb{R}^{3N}}\frac{\Delta E_{p}}{\Delta V}\rho_{\mathcal{H}}(u)du=\frac{2}{3V}\int_{\mathbb{R}^{3N}} E_{p}\rho_{\mathcal{H}}(u)du+\frac{1}{6V}\int_{\mathbb{R}^{3N}}\sum_{ijk}\Psi_{ijk}r_{i}^{eq}u_{j}u_{k}\rho_{\mathcal{H}}(u)du\label{eq:QHA}
\end{equation}
\end{strip}

The second term on the right-hand side of equation \ref{eq:QHA} corresponds
to the partial derivative of the free energy with respect to volume
in the quasiharmonic approximation. The first term on the right hand
side can be identified as a kinetic term using the quantum virial
theorem, which is missing if only the mean stress tensor from DFT is
taken, and reduces to the ideal gas expression in the limit of high
temperature. This result can be generalized to the case of an
anharmonic and anisotropic potential; marking the Cartesian directions $\alpha,\beta$
and the atomic index $i$, we write (as in Ref. \citep{SSCHA_stress_tensor}):

\begin{equation}
\sigma^{\alpha\beta}=\langle\sigma_{DFT}^{\alpha\beta}\rangle-\frac{1}{2V}\langle\sum_{i}u_{i}^{\alpha}f_{i}^{\beta}+u_{i}^{\beta}f_{i}^{\alpha}\rangle
\end{equation}

The present version of the code only handles the diagonal elements
of this stress tensor and thermal expansion is implemented for lattice
vectors along the cartesian directions or cubic lattices. In practice,
lattice parameters are iteratively modified until the calculated internal
pressure equals the target pressure within a chosen tolerance (of
the order of kbar).

\subsection{Reweighting, mixing, and handling imaginary frequencies}

As introduced for the SSCHA, one can reuse previously computed configurations
with the help of a reweighting scheme. Each configuration $u$
drawn at cycle $B$ thus gets an associated weight $w$ that depends
on the probability it would have to be drawn at the current cycle
$C$ based on the current thermal displacement matrix $\Sigma_{C}$
and on the probability it had to be drawn at cycle $B$ based on the
matrix that was used to draw it $\Sigma_{B}$:

\begin{equation}
w(u)=\frac{P_{\Sigma_{C}}(u)}{P_{\Sigma_{B}}(u)}
\end{equation}

This reweighting scheme is implemented using the memory parameter, such that all configurations of cycles $\left\lfloor C\left(1-\text{memory}\right)\right\rfloor$
to $C$ are taken into account. We point out that this scheme is formally
valid only at constant volume, such that in principle it should not
be used along with thermal expansion updates -- although in practice a
short memory can still be used, at the risk of slightly inaccurate
converged values.

How to handle the imaginary frequencies that might show up in the intermediate
phonon spectra is a problem that dates back to the original implementation
of the SCAILD method. At the time, it was chosen to switch them to
the real axis with the same modulus \citep{Souvatzis_SCAILD}. This
option is implemented in the code, but we also suggest another way
to deal with those frequencies, which is to arbitrarily fix them to
a finite real value. Those options are available using the imaginary\_freq
parameter.

Finally, it is usually better to use a strong mixing of the interatomic force
constants to improve the stability of the self-consistency cycle,
in particular close to phase transitions, since small stochastic deviations
might lead to strong divergences. We point out that this treatment is only a way to attain the fixed point of the algorithm in certain cases. The obtained phonon spectrum has physical meaning only if it is fully self-consistent, otherwise the persistent presence of imaginary frequencies or the inability to find a fixed point only indicates a strong likelihood of mechanical instability.

\subsection{Symmetries and acoustic sum rules}

In a first step, irreducible elements of the 2nd and 3rd order interatomic
force constants are identified using the crystal symmetries. Within
the matrix space that can be constructed from those irreducible elements,
we then identify eigenvectors that fulfill the acoustic sum rules
using singular value decomposition, so that any force constants matrix generated
by a combination of those eigenvectors respects the acoustic sum rule
by construction. The operations that generate the full force constants
matrix from each of those final irreducible elements are then stored
in sparse format. They are characteristic of the crystal symmetry,
so they can be used for compounds with the same structure but different
chemistry. This part of the code does not run in parallel at present.

\section{Example of SrTiO3}

\subsection{Convergence with respect to supercell size and wavevector grid}

\begin{figure}
\begin{centering}
\includegraphics[width=9cm]{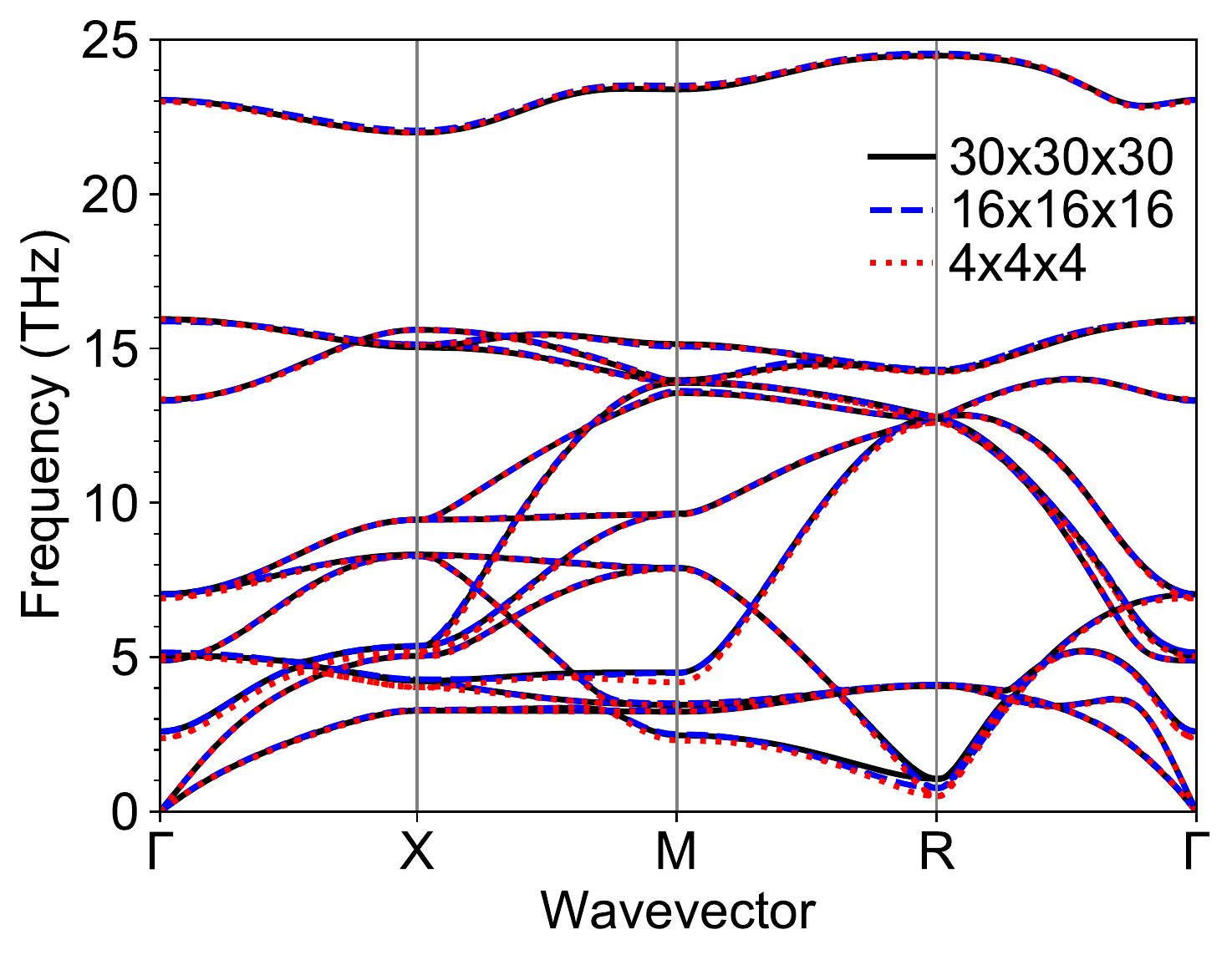}
\par\end{centering}
\caption{Band dispersion of cubic SrTiO$_{3}$ at 300\ K obtained with different samplings of the Brillouin zone for phonon properties, computed in a 4x4x4 supercell with $\Gamma$-point sampling for the electrons and the PBEsol functional.\label{band_grid}}

\end{figure}

\begin{figure}
\begin{centering}
\includegraphics[width=9cm]{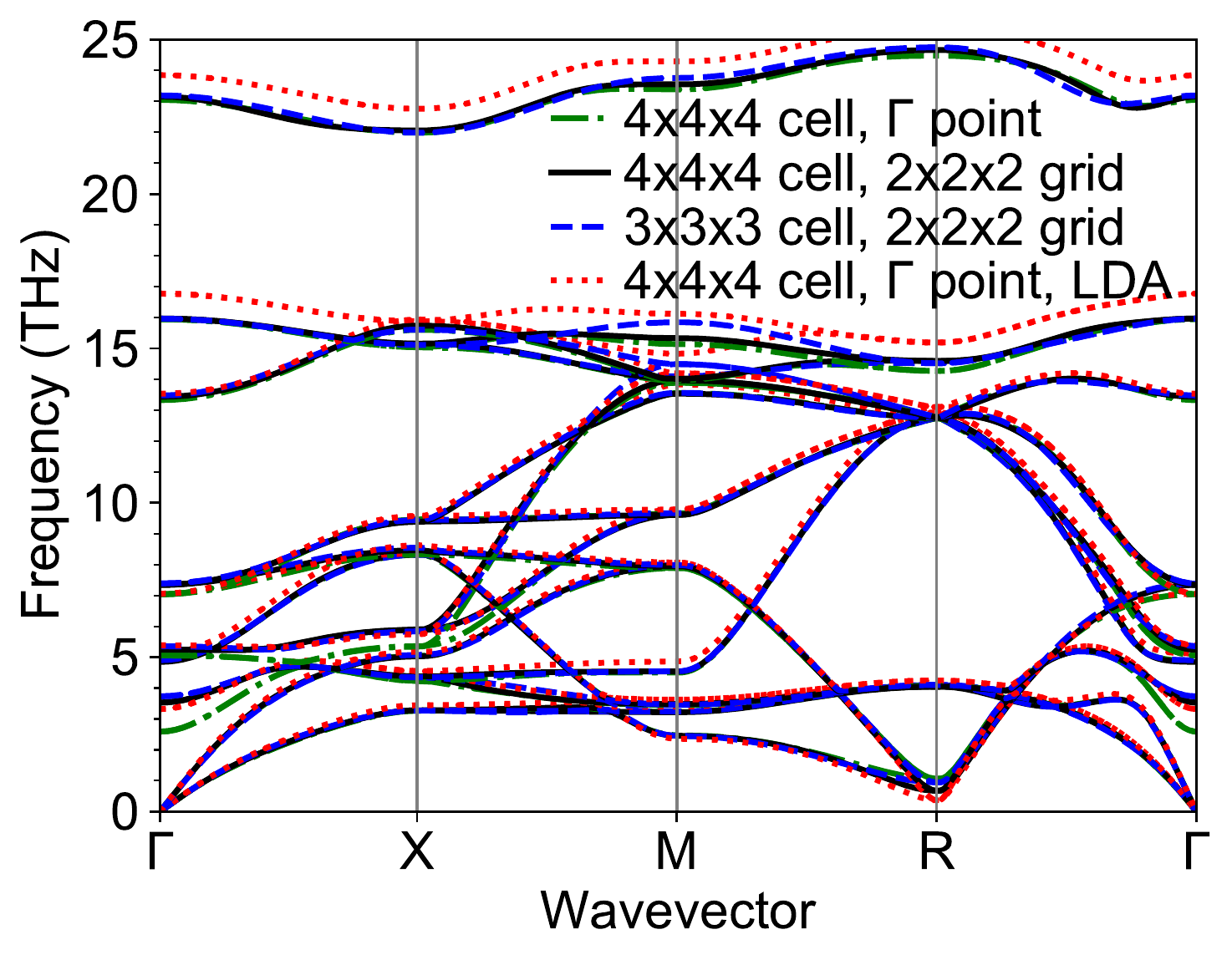}
\par\end{centering}
\caption{Band dispersion of cubic SrTiO$_{3}$ at 300\ K obtained with different supercells and samplings of the Brillouin zone for the electronic properties, using the PBEsol and LDA functionals.\label{band_supercell}}

\end{figure}

We computed the phonon dispersions and equilibrium lattice parameters
at 300 K for different supercells and wavevector grid sizes, with
convergence thresholds of $\pm0.2\;\mathrm{GPa}$ for the internal pressure and $\pm0.005\;\mathrm{eV/A^2}$ for the interatomic force constants. The obtained spectra are
displayed in Figs.~\ref{band_grid}~and~\ref{band_supercell}. The exchange-correlation functional was set to PBEsol except for a LDA calculation that is shown for the purposes of comparison. The non-analytical correction is applied based on ground-state Born charges and ion-clamped dielectric constant, but their temperature dependence could also be taken into account self-consistently \citep{Ambroise_dielectric}.  One can see that the dispersion converges rather quickly within about
0.5 THz, and that the use of a different exchange-correlation functional
appears to have a stronger impact. All in all, we typically observe a spread of the phonon frequencies between different calculations that translates into a spread of estimated transition temperatures of the order of 50\ K to 100\ K.

\subsection{Temperature dependence and typical errors}

\begin{figure}
\begin{centering}
\includegraphics[width=9cm]{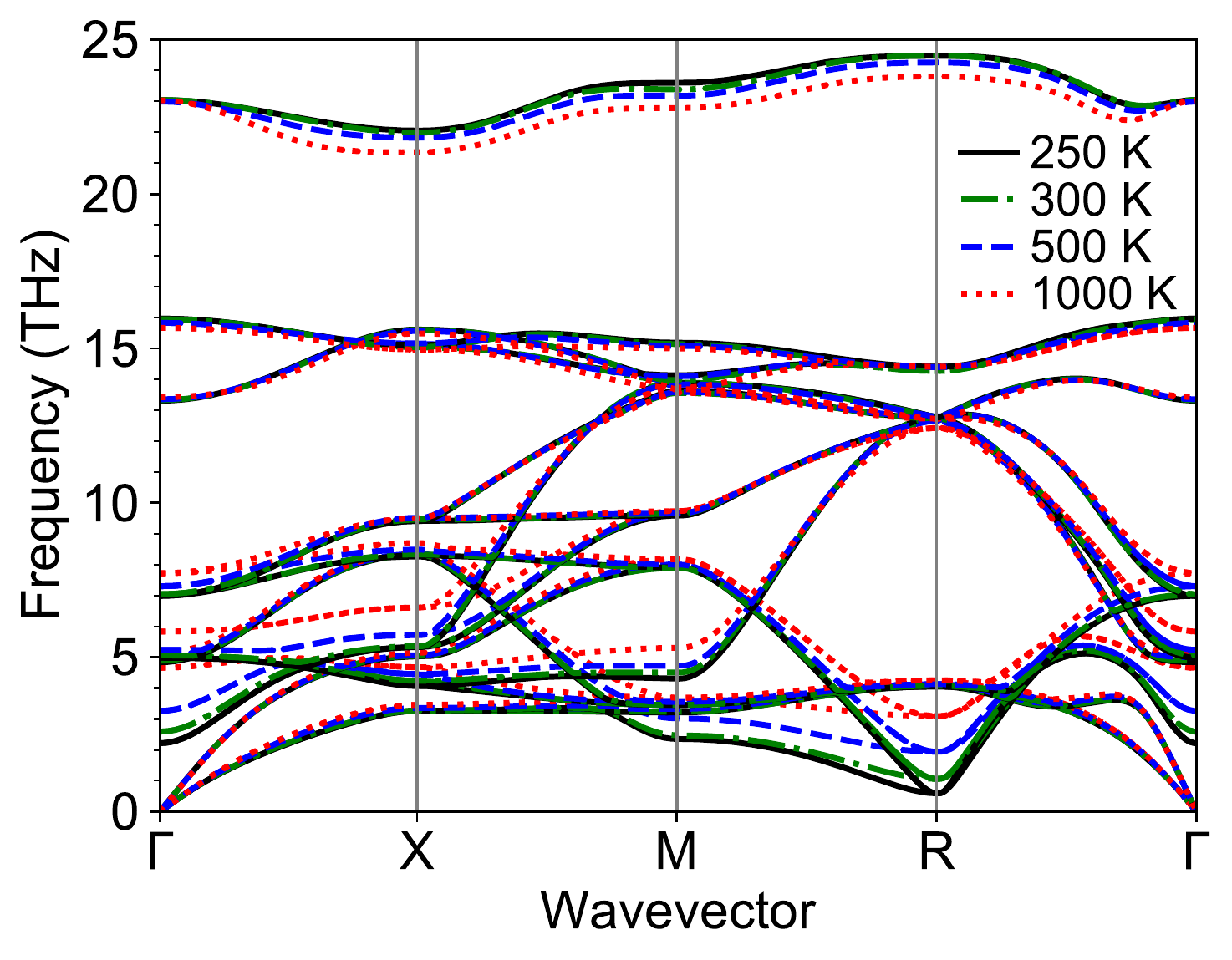}
\par\end{centering}
\caption{Temperature dependence of the phonon dispersion of cubic SrTiO$_{3}$ as obtained in the QSCAILD approach using the PBEsol exchange and correlation functional. \label{band_temperature}}

\end{figure}

\begin{figure}
\begin{centering}
\includegraphics[width=9cm]{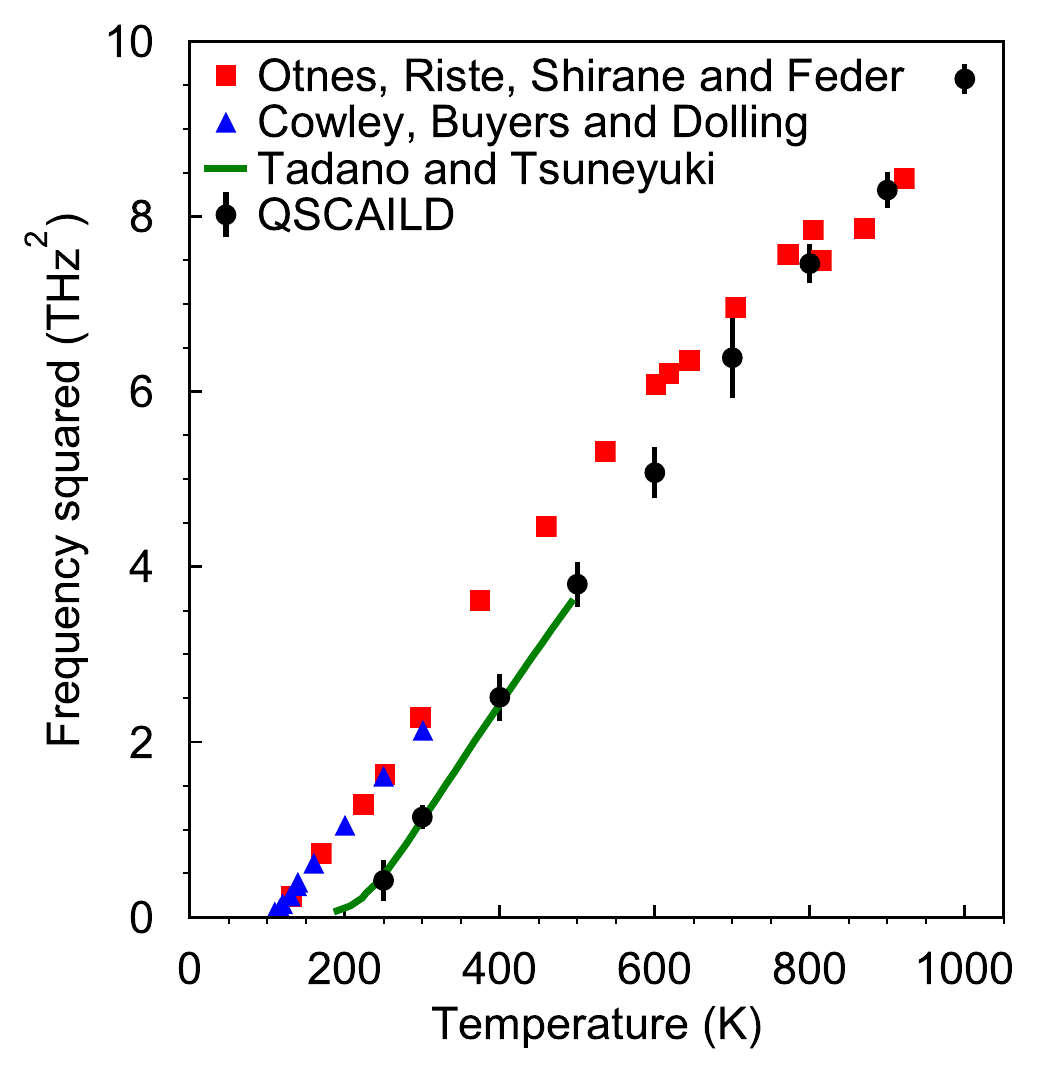}
\par\end{centering}
\caption{Temperature dependence of the square of the frequency of the soft mode at the R point of the Brillouin zone of cubic SrTiO$_{3}$. Experimental data from Refs.~\citep{Cowley_SrTiO3_temperature,Otnes_SrTiO3_temperature} and computed data from Ref.~\citep{Tadano_SCPH} are superposed for comparison. Average values and standard deviations of the frequency squared were estimated based on 5 converged calculations at each temperature.\label{soft_mode}}

\end{figure}

In the following, calculations were performed with a $4\times4\times4$ supercell
and a $30\times30\times30$ wavevector grid. In Fig.~\ref{band_temperature} we display
the phonon
dispersion of SrTiO$_{3}$ at different temperatures obtained with the PBEsol
exchange-correlation functional. Thermal expansion is taken into account, with a threshold for the average absolute pressure of 1\ kbar. The temperature dependence of the
soft mode displays the typical Curie-Weiss behavior (see Fig.\ \ref{soft_mode}),
which allows us to extrapolate the transition temperature to about 200\ K. We also note that it is in good agreement with the results obtained by Tadano and Tsuneyuki with the same functional \citep{Tadano_SCPH}. The average frequency squared values and standard deviations were estimated by restarting new calculations from the converged solutions, to obtain a total of 5 solutions for each temperature. This also shows a typical spread of the phonon frequencies that can be interpreted in terms of temperature as a precision of the method of about 50\ K to 100\ K, for this particular system and convergence criteria.

\section{Conclusion}

We have described and implemented a method to obtain
temperature-dependent,  second and third order effective interatomic
force constants. The method uses a regression analysis of forces from density
functional theory, coupled with a harmonic model of the quantum canonical
ensemble, and performed iteratively to achieve self-consistency
of the phonon spectrum. We have discussed the relationship of this technique
to other methods, showing that the resulting model minimizes the
free energy of the system within the assumption of a harmonic density
matrix, and proposed a way to include pressure from kinetic energy within this
static scheme. Finally, we have computed the temperature-dependent
force constants of SrTiO$_3$ with this method and found that it compares
well with experimental results, and with previous calculations using a different technique. The program
is open-source and available to the scientific community.

\section*{Acknowledgments}

This work was supported by HPC resources from GENCI-TGCC (project A0060910242).

\bibliographystyle{elsarticle-num}

\end{document}